%% file: main.tex
\documentclass[a4paper,11pt]{article}
\usepackage{jinstpub} 
\usepackage{color}
\usepackage[draft]{changes}
\definechangesauthor[name={lyh}, color=red]{lyh}
\definecolor{darkgreen}{rgb}{0.0,0.4,0.0}
\definechangesauthor[name={lsj}, color=darkgreen]{lsj}
\usepackage{appendix}
\usepackage{float}

\title{Reconstruction of Shower-like Events in NEON Using Likelihood and Graph Neural Network Methods}

\author[a,b]{Sujie Lin, }
\author[c,1]{Weiqin Huang,\note{Co-first author.}}
\author[c,2]{Yihan Liu,}
\author[c,3]{Chengyu Shao,}
\author[c,4]{Lili Yang, }
\author[c]{Huiming Zhang}

\affiliation[a]{State Key Laboratory of Physical Oceanography, Ocean University of China, Qingdao 266100, China}
\affiliation[b]{Joint Laboratory for Frontier Interdisciplinary Research in Ocean Science, Mathematics and Physics, Ocean University of China, Qingdao 266100, China}
\affiliation[c]{School of Physics and Astronomy, Sun Yat-sen University, No.2 Daxue Rd, 519082, Zhuhai China}
\emailAdd{liuyh363@mail.sysu.edu.cn}
\emailAdd{yanglli5@mail.sysu.edu.cn}
\emailAdd{shaochy@mail2.sysu.edu.cn}

\abstract{The Neutrino Observatory in the Nanhai (NEON) is a proposed deep-sea neutrino telescope deployed in the South China Sea. Accurate reconstruction of shower-like events is crucial for neutrino energy measurements and multi-messenger astronomy, yet it poses significant challenges due to seawater optical attenuation, irregular detector geometry, and substantial $^{40}\mathrm{K}$ ambient background. In this work, we present the first comprehensive reconstruction framework for shower-like events in NEON, encompassing both a physics-driven maximum likelihood estimation (MLE) method and a data-driven Graph Neural Network (GNN). The traditional MLE framework integrates spatial-isochronic hit selection, vertex reconstruction via time-residual M-estimator minimization, and decoupled directional and energy estimation based on pre-computed photon distribution tables. Physical calibrations, including PMT angular acceptance, hit-level time slewing corrections, and an effective line-source shower extension, are incorporated into the likelihood formulation. In parallel, a two-stage GNN is developed to capture intra-DOM PMT correlations and distance-weighted inter-DOM topological patterns. Simulation studies show that the MLE method achieves an overall median angular resolution of $4.19^\circ$ and an energy resolution of 25\%-37\% over 1 TeV to 1 PeV with negligible systematic bias. The GNN further improves reconstruction fidelity in the low-to-intermediate energy regime, achieving a median angular resolution of $1.8^\circ$ at 30 TeV and an energy resolution of $\sim$ 20\% between 40 and 300 TeV. Based on these reconstruction performances, the effective area and point-source discovery potential of NEON are evaluated. This framework establishes an essential reconstruction benchmark for NEON and provides practical methodologies for future next-generation deep-sea neutrino telescopes.}


\begin{document}

\maketitle
\flushbottom

\newcommand{\kn}{$^{40}\mathrm{K}$}

\section{Introduction} \label{sec:intro}
Since the first detection of astrophysical neutrinos in 2013 \cite{IceCube:2013cdw}, neutrino astronomy has opened a new window to high-energy phenomena in the Universe. Observations by IceCube and KM3NeT have revealed neutrino emissions associated with potential astrophysical sources, providing important insights into the origin of cosmic rays and multi-messenger astrophysics \cite{IceCube:TXS-2018, IceCube:NGC1068-2022, KM3NeT:2025npi}. However, identifying neutrino sources with high statistical significance and revealing diffuse flux spectral features remain major challenges, constrained by limited event statistics and reconstruction uncertainties.

Large-scale optical Cherenkov neutrino telescopes detect high-energy neutrinos through secondary relativistic charged particles produced in neutrino-nucleon deep inelastic scattering within transparent media (water or ice). These charged secondaries emit Cherenkov light recorded as discrete hit timing and charge measurements by arrays of photomultiplier tubes (PMTs). Accurately inferring the primary neutrino properties, such as arrival direction, deposited energy, and interaction vertex, constitutes the core computational task in neutrino experiments. The reconstruction quality affects point-source discovery sensitivity, energy spectrum, and prompt multi-messenger follow-up alerts. 

The Neutrino Observatory in the Nanhai (NEON) \cite{zhang2025proposed, Xie:2025njs} is a proposed next-generation deep-sea neutrino telescope deployed in the South China Sea, with an instrumented volume of approximately $10\text{ km}^3$. The array comprises vertical detector strings arranged in a Fibonacci spiral configuration \cite{mungubaComplexBuildAlgorithm2021}, with each string equipped with over ten digital optical modules (DOMs) containing densely packed PMTs. Compared to regular grid arrays, NEON is characterized by a sparse, irregular spatial geometry coupled with dynamic deep-sea optical properties, presenting unique demands on event reconstruction algorithms.

High-energy neutrino-nucleon interactions typically manifest as track-like or shower-like topologies. Muon neutrino charged-current ($\nu_\mu$ CC) interactions produce muons traveling kilometer-scale distances, yielding elongated ``tracks'' well-suited for directional reconstruction. Conversely, electron neutrino charged-current ($\nu_e$ CC) interactions generate compact, localized electromagnetic and hadronic cascades with characteristic longitudinal extensions of $\mathcal{O}(10\text{ m})$. In this paper, our reconstruction methods are evaluated and benchmarked primarily using $\nu_e$ CC events. Nevertheless, because neutral-current (NC) interactions produce an essentially identical compact cascade topology, the proposed framework can be directly applied to NC events as well.

Despite their compact morphology, shower-like events provide unique advantages in neutrino astrophysics and particle physics \cite{Barger:2013pla, Sudoh:2023qrz}. The complete calorimetric containment of cascade energy enables precise, unbiased reconstruction of the parent neutrino energy, making cascades particularly sensitive to spectral breaks and the $\sim 6.3\text{ PeV}$ Glashow resonance ($\bar{\nu}_e + e^- \to W^-$)\cite{IceCube:2021rpz}. Furthermore, the substantial suppression of atmospheric muon backgrounds in contained cascade selections permits relaxed fiducial cuts, significantly expanding the effective detection volume for rare astrophysical neutrino fluxes.

However, accurately reconstructing shower-like events is quite difficult. Secondary relativistic electrons undergo severe multiple Coulomb scattering, rapidly deflecting and randomizing their trajectories. Consequently, Cherenkov emission from cascades forms a quasi-isotropic, diffuse light distribution that lacks the clear spatial-temporal wavefront progression of muon tracks. In addition, seawater optical scattering and absorption disperse photon arrival times, while random coincidence hits from ambient $^{40}\text{K}$ decays severely degrade signal-to-noise ratios, especially at low-to-intermediate energies.

There are two reconstruction algorithms in the field for high-energy neutrinos, physics-driven maximum likelihood estimation (MLE), and data-driven machine learning. Traditional MLE frameworks typically adopt a multi-stage approach, where interaction vertices are first localized by minimizing arrival time residuals under spherical wavefront assumptions, followed by directional and energy optimization using pre-computed photon probability tables (as successfully implemented in KM3NeT \cite{vanEeden:2021zzv}). On the machine learning frontier, while Convolutional Neural Networks (CNNs) have proven effective for regular, pixelated detector geometries (e.g., IceCube \cite{IceCube:2025jmv}), the irregular and sparse point-cloud structure of NEON events is naturally suited to Graph Neural Networks (GNNs) \cite{abbasi2022graph, reck2021graph}.

In this work, we present the first comprehensive reconstruction framework for shower-like events in NEON, including both a calibrated likelihood-based pipeline and a hierarchical GNN architecture. The simulation chain, encompassing neutrino fluxes, deep-inelastic scattering, and full optical propagation in deep-sea water, follows our established framework (refer to \cite{zhang2025proposed} for details). For detector digitization, an ambient $^{40}\text{K}$ noise rate of $110\text{ kHz}$ per DOM and a basic discriminator threshold of $0.3$ photoelectrons (PE) are implemented. The baseline NEON geometry consists of 1,200 strings deployed over a 10-km diameter seafloor region, modularized into seven identical circular sub-arrays (radius $1.79\text{ km}$) to optimize trigger efficiency and minimize undetectable signal losses.

This paper is organized as follows. Section~\ref{sec:trad_algo} describes the likelihood-based reconstruction method. Section~\ref{sec:gnn_algo} presents the GNN-based approach. Section~\ref{sec:performance} discusses the reconstruction performance. Finally, conclusions are given in Section Section~\ref{sec:conclu}.

\input{sec_trad_algori}

\section{Graph Neural Network Reconstruction} \label{sec:gnn_algo}


GNNs are a type of neural network architecture well-suited for processing point cloud data. NEON's shower-like events (Cherenkov signals produced after neutrino collisions) naturally forms a point cloud, which can be effectively represented as a graph with nodes and edges. 
Each observed pulse is treated as a node with features 
including its 3D position, the recorded hit time, and the measured charge. These nodes inherently possess relational structure. Connections
are established between each node and several of its nearest neighbors based on Euclidean distance. The number of neighbors can be adjusted according to specific requirements. Due to the spatial distribution of the detectors, the data effectively form an irregular point cloud. To address this irregularity, we have designed a dedicated two-stage aggregation model.

\subsection{Event Preprocessing}

Due to the presence of biological luminescence and \kn\ decay in the deep-sea environment, substantial background noise is introduced into the data. Including all such noise in the GNN would introduce significant interference and considerably reduce computational efficiency. Therefore, we perform noise-filtering preprocessing as mentioned in the previous section.

To facilitate stable and efficient training of the GNN, we apply normalization preprocessing to the input features. Instead of using distribution-dependent statistics (e.g., per-feature mean and standard deviation), which can be sensitive to outliers and may distort the inherent physical relationships within an event, we employ a simple, uniform scaling factor of 1/1000. This converts positional coordinates from meters to kilometers and timing from nanoseconds to microseconds, preserving the relative geometric and temporal scales of the shower development within the detector volume. This approach is robust, deterministic, and agnostic to the overall data distribution, ensuring consistent treatment for all events.


After noise filtering, the dataset comprises approximately 35,000 events, which we split into training, validation, and test sets in an 8:1:1 ratio. To ensure sufficient training volume, we perform data augmentation by expanding the training set through rotation and translation transformations. These operations increase the sample size while preserving the underlying physical correlations. Augmentation is applied exclusively to the training set to maintain the integrity of the validation and test sets, ensuring an unbiased evaluation of the model's generalization performance. This strategy enables robust learning from a limited dataset.

\subsection{Model Architecture}

In previous studies \cite{abbasi2022graph,reck2021graph}, GNNs have typically been applied by treating each individual PE hit as a node, with neighbors determined through spatial proximity for message passing. However, directly applying this approach to our data presents a challenge due to the heterogeneous spatial distribution of PMTs. Those within a single DOM are positioned closely together, whereas distances between PMTs across different DOMs are significantly larger.

In shower-like events, PMTs within the same DOM exhibit strong intrinsic correlations. The collective hits on a single DOM define its key features, such as the temporal and spatial distribution of photons and the total PE count, which reflect the local direction and intensity of incident Cherenkov light. In contrast, the hit pattern across different DOMs follows the collective radiation profile expected from the global shower morphology. To accommodate this hierarchical structure, we adopt a two-stage graph architecture. The first stage captures correlations among PMTs within the same DOM, while the second stage captures patterns across different DOMs.

\begin{figure}
\centering
\includegraphics[width=0.25\linewidth]{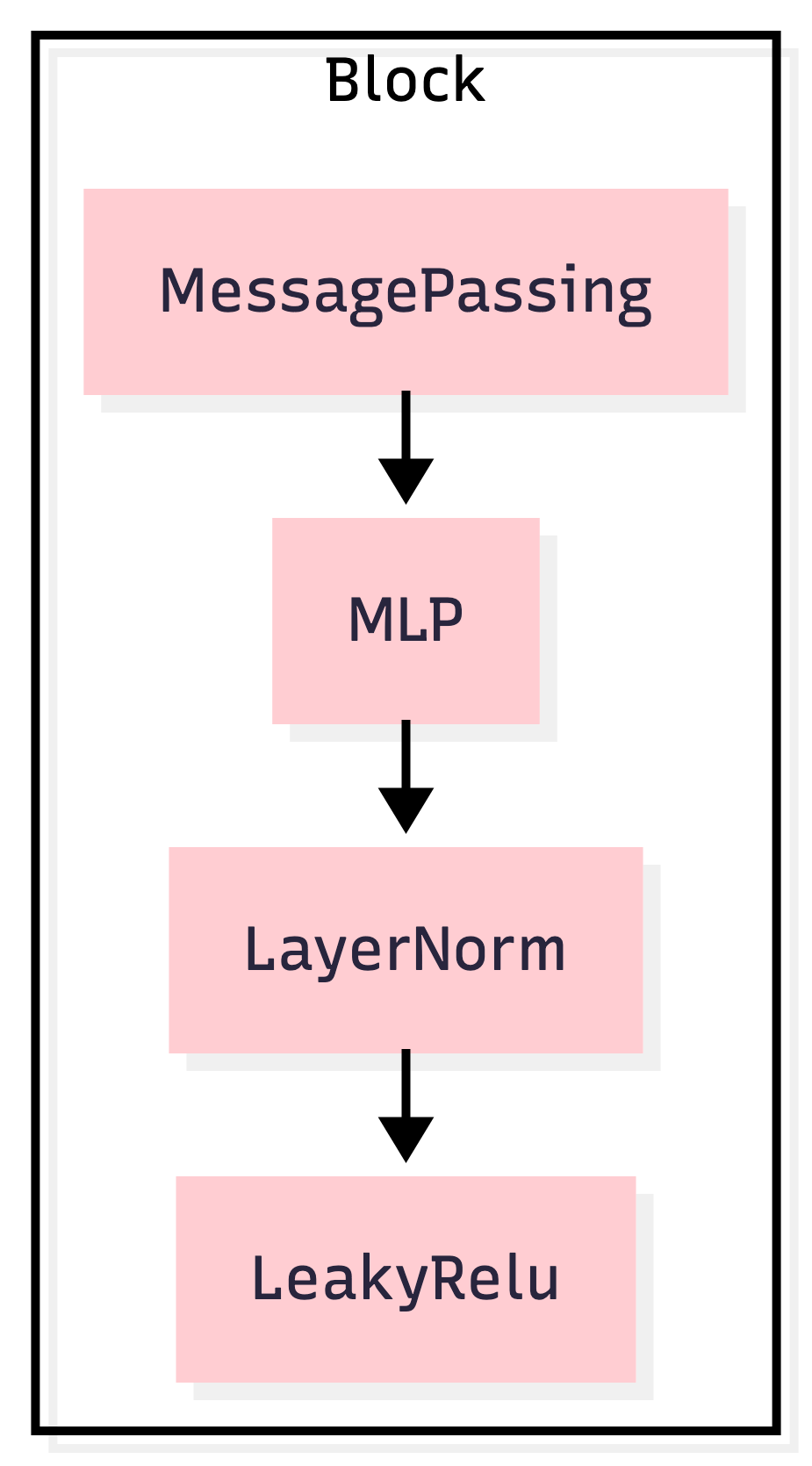}\hfill
\includegraphics[width=0.35\linewidth]{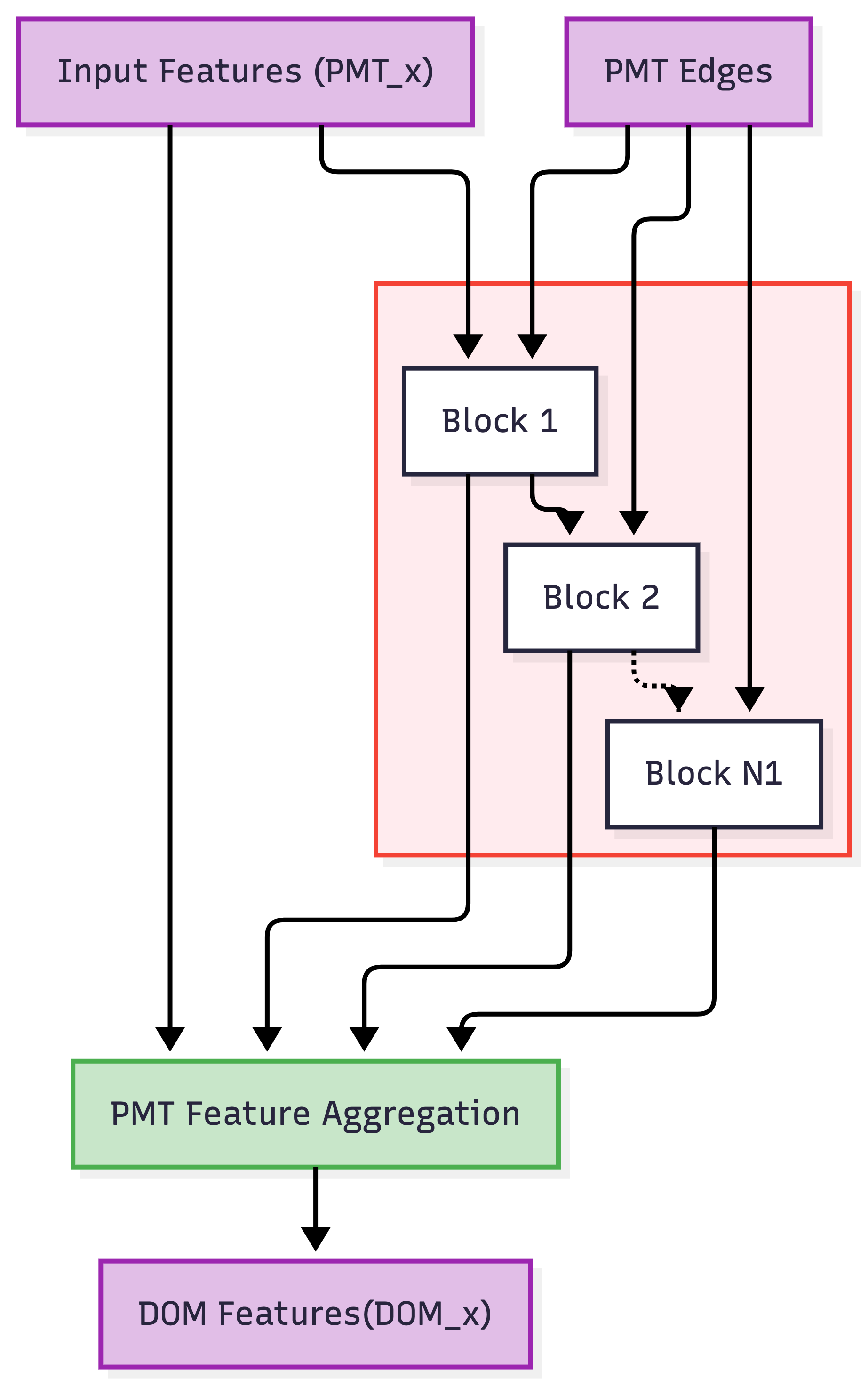}\hfill
\includegraphics[width=0.35\linewidth]{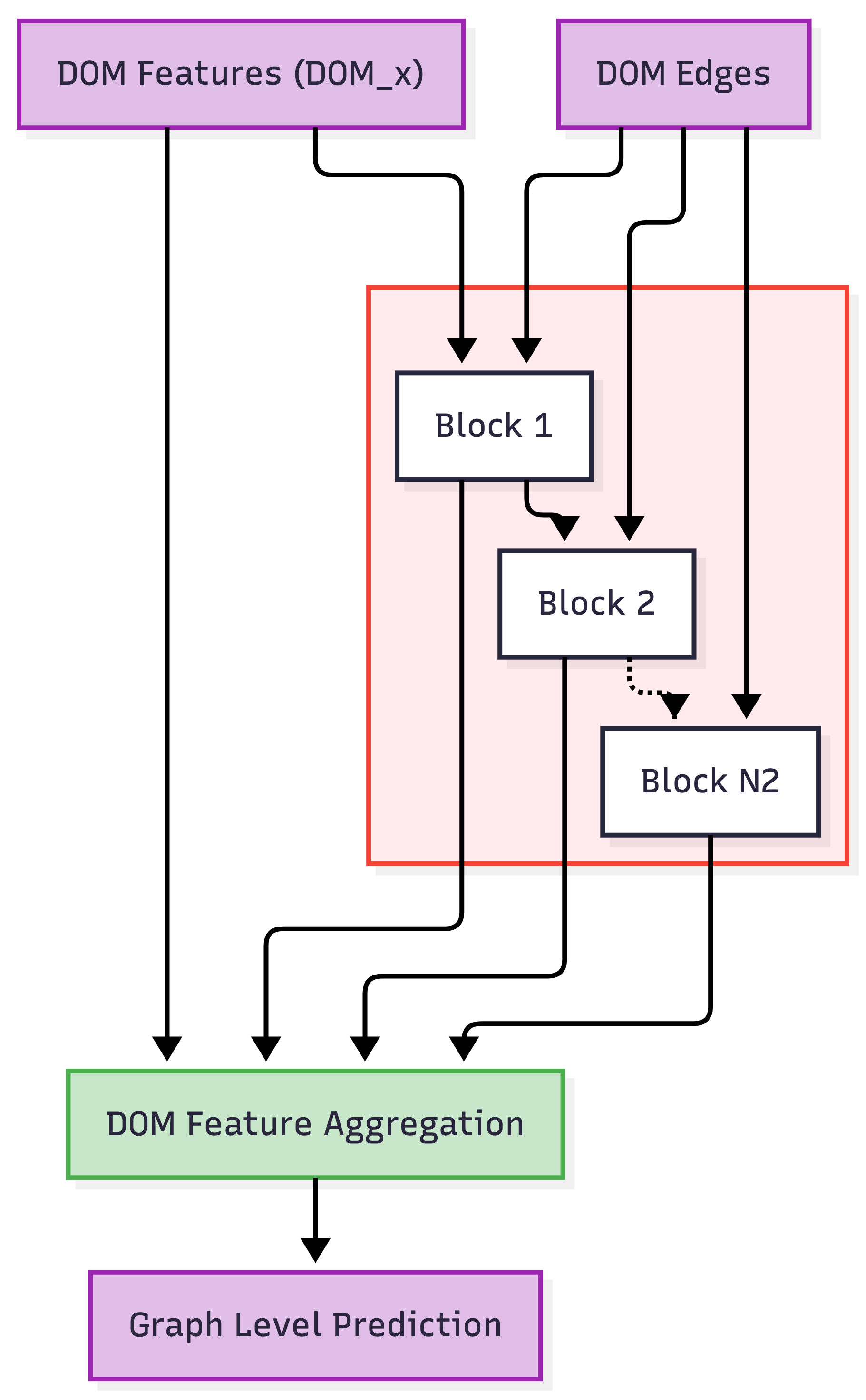}
\caption{Model architecture components. Left: internal structure of a single MessagePassing Block, composed of an MLP, a normalisation layer, and a LeakyReLU activation layer.
Middle: PMT-level processing stream, where PMT features inside each DOM are aggregated via intra-DOM PMT edges and iteratively refined through $N_1$ MessagePassing Blocks.
Right: DOM-level processing stream, where DOM features are aggregated via inter-DOM edges and iteratively refined through $N_2$ MessagePassing Blocks.}
\label{fig:flow}
\end{figure}

As shown in Figure~\ref{fig:flow}, our model begins with the PMT features (PMT\_x) and processes them through a sequence of $N_1$ computational units termed Blocks. Each Block consists of a message-passing step, a multilayer perceptron (MLP), layer normalization, and a LeakyReLU activation. The message-passing operation within these PMT-level Blocks is designed to capture intra-DOM relationships and is performed as follows: for each PMT node $j$, we aggregate information from its neighboring PMTs $i$ within the same DOM according to
\begin{equation}
\label{eq:intra_om_msg}
x_j = \sum_{i \in \mathcal{N}_{\text{intra}}(j)} \left( x_j,\, x_j - x_i \right).
\end{equation}
After the PMT features have been refined through $N_1$ such Blocks, they are pooled per DOM to form a single DOM-level feature, which becomes the node representation for the subsequent DOM graph.

The DOM-level features (DOM\_x) then undergo a similar transformation through another stack of $N_2$ Blocks. While each Block's internal structure remains unchanged, the message-passing step now operates over an inter-DOM graph whose edges connect DOMs within a 150 m spatial radius. This inter-DOM message passing incorporates an inverse-distance weighting scheme to modulate the influence of neighboring nodes. For a DOM node $j$, the weighted aggregation is given by
\begin{equation}
\label{eq:inter_om_agg}
x_j = \sum_{i \in \mathcal{N}_{\text{inter}}(j)} w_{ij} \, \left( x_j,\, x_j - x_i \right), \qquad w_{ij} \propto \frac{1}{d_{ij}},
\end{equation}
where $d_{ij}$ is the spatial distance between DOMs $i$ and $j$. This weighting emphasizes nearby DOMs while suppressing potential noise from more distant modules.

To preserve hierarchical information across the processing stages, we retain the output features from every Block in both processing streams. These features are then aggregated using three parallel pooling operations: mean, min, and max.

All relevant model parameters, including structural hyperparameters such as the number of Blocks ($N_1$, $N_2$) and layer widths, as well as training hyperparameters like the learning rate, are optimized using the Optuna framework, which performs an efficient Bayesian search to identify the configuration that maximizes validation performance \cite{Akiba:2019lwq}.

\begin{figure}
\centering
\includegraphics[width=0.6\linewidth]{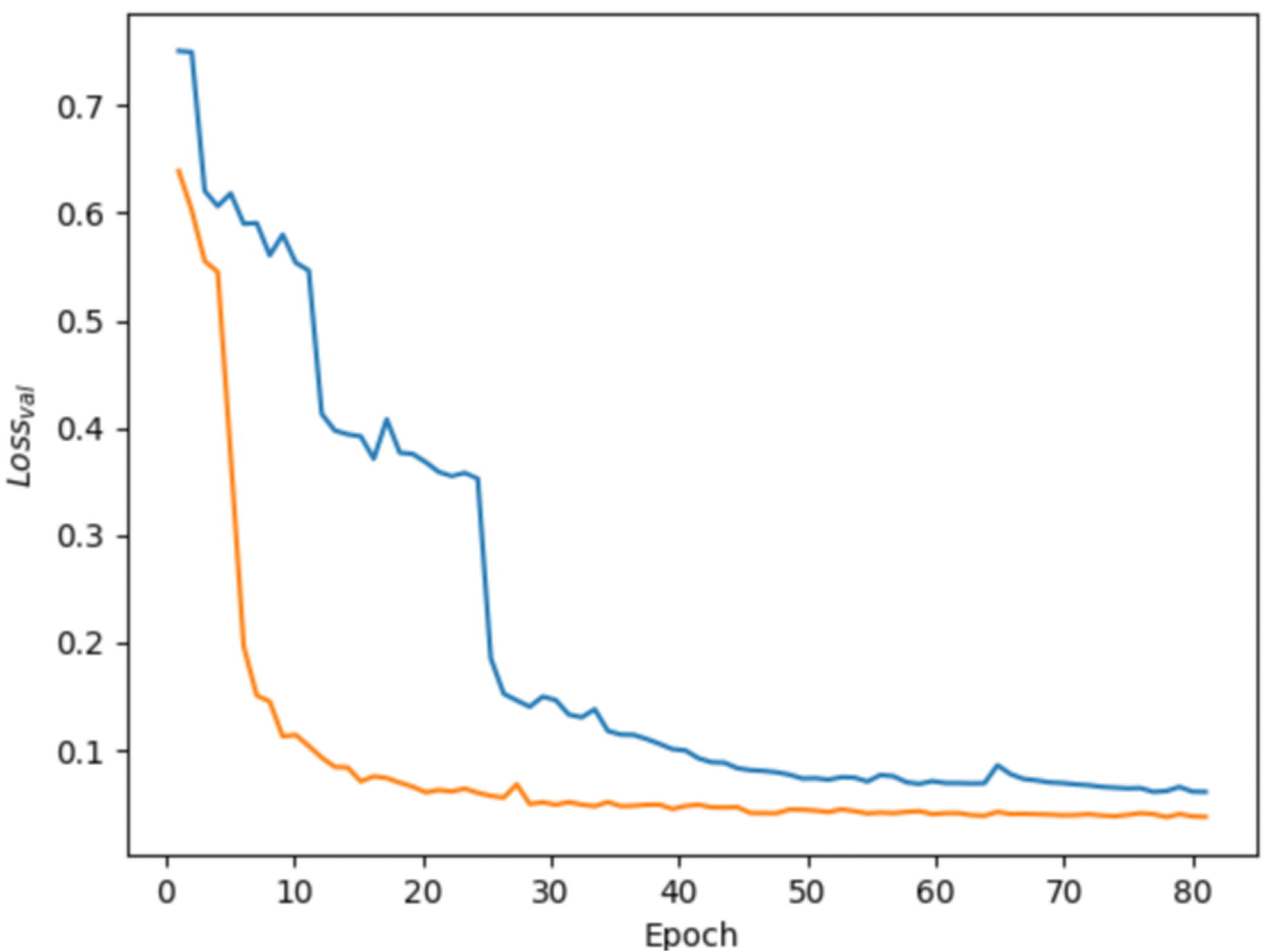}
\caption{Training loss curves.
The blue curve shows the baseline two-stage architecture without
inter-DOM weighting. Its loss decreases in pronounced stages and reaches the ${\sim}0.1$ level only after about 25 epochs.
The orange curve corresponds to the proposed model that incorporates
distance-weighted inter-DOM message passing and radius-based neighbor
selection; the loss descends rapidly and reaches ${\sim}0.1$
within 10 epochs, with smoother convergence throughout.}
\label{fig:losscurve}
\end{figure}

As illustrated in Figure ~\ref{fig:losscurve}, this proposed two-layer GNN architecture, incorporating distance-weighted message passing and radius-based neighbor selection, shows superior training dynamics compared to the baseline. Specifically, the new model exhibits a faster decrease in training loss, smoother convergence with reduced curve fluctuations, and ultimately reaches a lower and more stable loss value upon convergence.

\input{sec_perform_disc}
\input{sec_conclu}

\acknowledgments

We thank Caijin Xie, Zijian Qiu and Yudong Cui for useful discussions. This work is supported by the National Natural Science Foundation of China (NSFC) grants 12261141691.

\input{appendix}


\bibliographystyle{JHEP}
\bibliography{biblio.bib}




\end{document}

%% file: sec_trad_algori.tex
\section{Likelihood-based Reconstruction of Cascade-like Events}\label{sec:trad_algo}

\subsection{Hit Selection and Preprocessing}

We develop an efficient logic for DOM triggering and hit selection that exploits the isochronic space–time distribution of signal hits to suppress background noise such as $^{40}$K photons. This relies on two assumptions. Secondary electrons have short/point-like tracks in seawater, and the speed of seabed Cherenkov photons cannot exceed the local light speed ($\frac{c}{n_\mathrm{w}}$, $n_\mathrm{w}$ as the refractive index of water).


The DOM trigger identifies which DOMs have recorded signals and applies an initial selection to filter out unlikely hits.
A DOM is triggered by either a large-PE pulse (e.g., $\mathrm{PE}>10$) or a dense hit cluster 
(e.g., more than 3 hits within $50\,\mathrm{ns}$), none of which is typical of background 
\kn\ noise. Hits occurring far earlier than the majority (e.g., beyond $5\sigma$ from the 
mean time) are rejected, and the earliest surviving hit is adopted as the DOM triggering hit. 
A readout window of $\Delta t$ (e.g., $-1000\,\mathrm{ns}$ 
to $+5000\,\mathrm{ns}$) relative to this time is then opened to collect candidate signal hits.

To recover genuine signal hits discarded by the initial trigger, especially at distant DOMs from the vertex and too weak hits to trigger.
We perform a spatial--isochronic reselection of signal hits firstly by a fast vertex search and then by including hits near the Cherenkov isochronic sphere. The isochronic residual for each hit is defined as
\begin{equation}
\Delta t_{\rm iso} = t_i - \left(t_0 + \frac{|d_{i-0}|}{c_w}\right) \gtrsim 0,
\end{equation}
where $t_i$, $t_0$, $d_{i-0}$, and $c_w$ are the hit time, vertex time, distance from the vertex, and light speed in seawater, respectively.
As shown in Figure~\ref{fig:pe_linearity}, this procedure efficiently separates signals from the \kn\ backgrounds. More signal hits ($\sim$92\%) are retained, while noise is suppressed to $\sim$0.2\% of the pre-selection level. Although further noise suppression within the time window is desirable, current performance already suffices for reconstruction.

\subsection{Vertex Reconstruction}

We first obtain a preliminary estimate of the vertex time and position by averaging the selected hits, following a procedure similar to that described above.
Previous studies have demonstrated that reconstructions that exploit the arrival times of hits on the isochronous sphere are robust and produce precise predictions \citep{2017ICRC...35..950M}.
The exact vertex is then determined by minimizing the weighted time residuals of the selected hits. We employ the NLOPT package \citep{NLOPT} with the GN\_ISRES global-optimization algorithm to minimize an M-estimator \citep{2017ICRC...35..950M},
\begin{equation}
\mathcal{M}_{\rm est} = \sum_{i\in hits} \Bigl( w_i \sqrt{1 + t_{{\rm res},i}^{2}} + p(t_{{\rm res},i}) \Bigr),
\end{equation}
where $t_{{\rm res},i} = t_i - \bigl(t_0 + |d_{i-0}|/c_w\bigr)$ is the time residual of the $i$-th hit, and $w_i$ is its PE weight. 
The penalty term $p(t_{\rm res})$ suppresses unphysical early hits ($t_{\rm res}<0$, penalized by $2000\,|t_{\rm res}|$) and heavily scattered late photons ($t_{\rm res}>10\,\mathrm{ns}$, penalized by $800\,|t_{\rm res}|$), while allowing a $10\,\mathrm{ns}$ tolerance window for detector resolution and finite shower size ($0 \leq t_{\rm res} \leq 10\,\mathrm{ns}$, $p=0$).
Early hits ($t_{\rm res}<0$) are therefore heavily penalized because photons cannot arrive earlier than the direct Cherenkov light front. A tolerance window of $0 \leq t_{\rm res} \leq 10$\,ns accommodates detector time resolution and the finite shower size. Larger positive residuals are moderately penalized as they likely arise from scattered photons.
Events are required to contain at least four triggered DOMs that are non-coplanar, ensuring that the isochronous sphere is geometrically constrained, which will be excluded otherwise.
The resulting vertex reconstruction achieves a mean spatial error of $6.5\,\mathrm{m}$ for cascade events with energies above $10\,\mathrm{TeV}$.

\subsection{Direction and Energy Reconstruction}
\label{sec:joint}

\begin{figure}[htbp]
    \centering
    \includegraphics[width=0.36\linewidth]{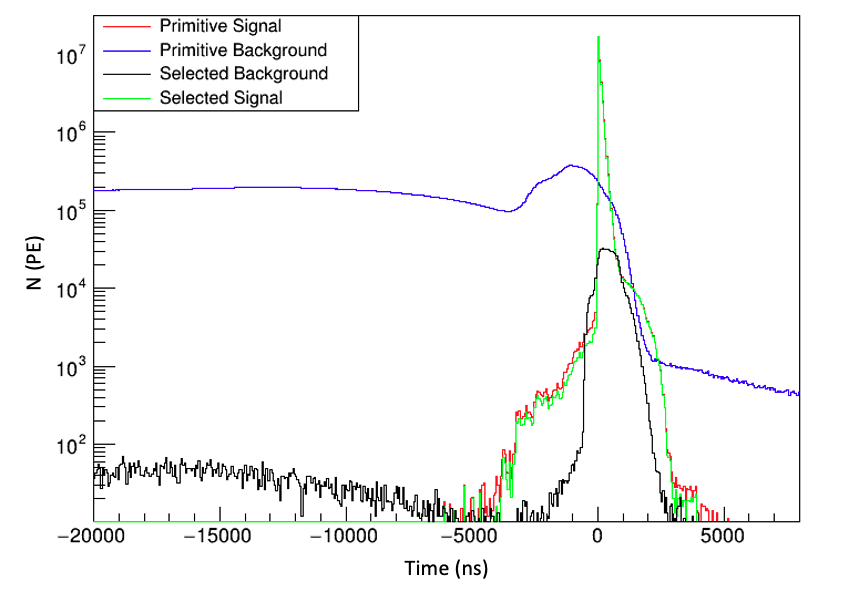}
    \includegraphics[width=0.6\linewidth]{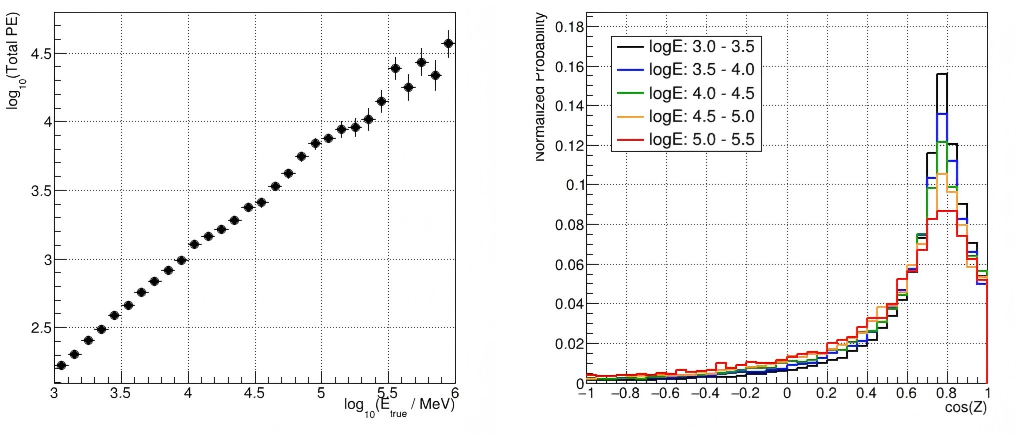}
    \caption{Statistics about simulated hits in shower-like neutrino detections. \textit{Left:} Hit (in PE) time distributions about the isochronic sphere surface. \textit{Middle: }
    Profile of total observed PE in the fiducial volume versus
    true neutrino energy.
    \textit{Right:} Normalized angular distributions of Cherenkov signals about the shower direction  in different energy ranges between 1\,TeV and 1\,PeV.}

    \label{fig:pe_linearity}
\end{figure}

Other studies reconstructed the direction and energy simultaneously \citep{2017ICRC...35..950M,Middell2009ImprovedRO}. 
Nevertheless, in this study, the high computational cost of generating high-energy cascade events results in a relatively small data sample, precluding a stable bin-by-bin determination of the energy-dependent PDFs, most critically for energies exceeding 100\,TeV. Moreover, the correlated energy and directional responses of the detector may compromise the fit via parameter degeneracy. Consequently, we decouple the two estimation procedures and handle them independently.
The total PE yield gives the energy directly, as it is approximately proportional to the shower size. The spatial Cherenkov profile is nearly self-similar from $1\,\mathrm{TeV}$ to $1\,\mathrm{PeV}$, so a single spatial PDF is sufficient for directional reconstruction.
Figure~\ref{fig:pe_linearity} shows this linearity and similarity. The middle panel presents the linear relationship between PE and true energy, confirming that total Cherenkov yield is proportional to deposited shower energy across the full simulated range, justifying the analytic energy estimator of Eq.~(\ref{eq:erec}). The right panel shows the Cherenkov light distribution with various energy.

 
\subsubsection{Likelihood Formulation}
\label{sec:likelihood}

The MLE for the shower direction is constructed by contrasting the observed Cherenkov hits against the expected ones. The shower is characterized by three probability distributions, the total photon yield $\Lambda$ (in PE), the normalized angular probability of photon hits $P_{\rm ang}$, and the time-residual probability distribution $P_{\rm time}$. Calibrations for PMT angular acceptance and signal time slewing are also incorporated, and details of these PDFs and calibrations are given in Appendix~\ref{app:mle}.

For a neutrino with direction $\mathbf{d}$, the expected PE on PMT $i$ is
\begin{equation}
    \mu_i = \Lambda(R_i) \cdot P_{\rm ang}(R_i,\cos\theta_i) \cdot w_{{\rm acc},i} \cdot S + \mu_{\rm bg},
\label{eq:expectedcharge}
\end{equation}
where $R_i$ is the vertex-to-PMT distance, $\cos\theta_i = \mathbf{d}\cdot\hat{\mathbf{r}}_i$ is the cosine of the angle between the shower axis and the unit vector from the vertex to PMT $i$, $w_{{\rm acc},i}$ is the angular acceptance function (Appendix~\ref{sec:calib}), $S$ is a global energy scaling factor (Section~\ref{sec:energy}), and $\mu_{\rm bg}$ is the uniform background PE. The background term $\mu_{\rm bg} = 5\times10^{-4}\,\mathrm{PE}$ per PMT per time window is obtained by clipping hits unlikely to be signals ($\gtrsim 3000$\,ns before/after a shower).

The total negative log-likelihood of an assumed shower direction $\hat{\mathbf{d}}$ is the sum of a spatial term over all PMTs and a temporal term over all individual hits,
\begin{equation}
    -\ln\mathcal{L}(\hat{\mathbf{d}}) = \sum_{i\in\mathrm{PMTs}} \mathcal{L}_{{\rm space},i} + \sum_{j\in\mathrm{hits}} \mathcal{L}_{{\rm time},j}.
\label{eq:nll}
\end{equation}

The spatial term compares observed PE $n_i$ to predicted $\mu_i$. For under-illuminated PMTs ($n_i \le \mu_i$) we use the exact Poisson likelihood, and for over-illuminated PMTs ($n_i > \mu_i$) we switch to a Gaussian approximation with variance $\sigma^2 = \mu_i + 0.04\mu_i^2$ (accounting for systematic charge fluctuations in large signals), and apply a linear tail beyond $3\sigma$ to limit outlier influence,
\begin{equation}
    \mathcal{L}_{{\rm space},i} =
    \begin{cases}
        \mu_i, & n_i = 0, \\[4pt]
        (\mu_i - n_i) - n_i\ln\!\dfrac{\mu_i}{n_i}, & 0 < n_i \le \mu_i, \\[6pt]
        \tfrac{1}{2}z^2 + \tfrac{1}{2}\ln\sigma^2, & n_i > \mu_i,\;z \le 3, \\[4pt]
        4.5 + 3(z-3) + \tfrac{1}{2}\ln\sigma^2, & n_i > \mu_i,\;z > 3,
    \end{cases}
\label{eq:spatial}
\end{equation}
where $z = (n_i - \mu_i)/\sigma$.

The temporal term penalizes deviations of each hit's arrival time from the expected Cherenkov propagation,
\begin{equation}
    \mathcal{L}_{{\rm time},j} = -\,q_j\,\ln P_{\rm time}(R_i,\,\cos\theta_i,\,t'_{{\rm res},j}).
\label{eq:temporal}
\end{equation}
For this segment we adopt a finite line-source model for the vertex (rather than a point source) to improve reconstruction performance. The calibrated time residual $t'_{{\rm res},j}$ and line-source details are given in Appendix~\ref{app:lsm}. Here, $R_i$ and $\theta_i$ are the hit distance and angle from the shower vertex as above; $P_\mathrm{time}$ denotes the time-residual probability distribution as in Appendix~\ref{sec:pdftables}.  Weighting by the individual hit PE $q_j$ gives larger hits more influence in the fit.

\subsubsection{Two-Stage Directional Optimization}
\label{sec:direction}
 
In practice, Negative Likelihood Landscape (NLL) surface in direction space is of multiple local minimums or even highly non-convex for shower-like events, because of weak directional constraints alone from the nearly isotropic Cherenkov signal and the background.
We address this with a two-stage strategy as follows.
 
\paragraph{Pass~1: Global Seed Search.}
We sample the full unit sphere with $N = 1000$ uniformly distributed
seed directions, generated by a Fibonacci (golden-ratio) spiral
lattice~\cite{Gonzalez2010,Hannay2004},
\begin{equation}
    \theta_k = \arccos\!\left(1 - \frac{2k+1}{N}\right)
    \qquad
    \phi_k   = \frac{2\pi k}{\varphi}\;\mathrm{mod}\;2\pi,
\label{eq:fibonacci}
\end{equation}
where $\varphi = (1+\sqrt{5})/2$ is the golden ratio and
$k = 0,\ldots,N-1$, providing near-optimal uniform spherical coverage even near the poles.
The NLL is evaluated at each seed with the energy scaling factor $S$
fixed at a reference value ($\log_{10}E_{\rm ref} = 4.0$, i.e.\
$10\,\mathrm{TeV}$), decoupling the direction search from energy
estimation.
The choice $N = 1000$ follows from a simple geometric argument that the average solid angle per seed is $4\pi/N \approx 0.013\,\mathrm{sr}$,
corresponding to a mean inter-seed angular spacing of about $6^{\circ}$.
The top-10 seeds by NLL are each refined by a local Constrained Optimization 
and the best result is taken as the
Pass~1 estimate.
 
\paragraph{Pass~2: Local Refinement.}
optimization constrained to the unit sphere
($\lVert\hat{\mathbf{d}}\rVert_2 = 1$), with initial step size
$10^{-2}\,\mathrm{rad}$ and convergence tolerance $10^{-5}$.
The step size $10^{-2}\,\mathrm{rad}$ was chosen by testing values
from $10^{-1}$ to $10^{-3}$. Smaller values caused the optimizer to
stall in the local curvature while larger values caused it to jump away
from the minimum.
 
\paragraph{Quality Cut: Seed Dispersion.}
After Pass~1, we compute the angular dispersion among the top-15
COBYLA results as a proxy for how well-constrained the likelihood
minimum is.
Events where this dispersion exceeds $10^{\circ}$ are rejected, as
they correspond to degenerate topologies where the shower is too faint
or too poorly sampled for reliable directional reconstruction.
The $10^{\circ}$ threshold was determined by scanning values from
$5^{\circ}$ to $20^{\circ}$ and choosing the point that maximized the
product of reconstruction efficiency and median angular resolution on
the validation set.
 
\subsubsection{Energy Reconstruction}
\label{sec:energy}
 
Once the optimal direction $\hat{\mathbf{d}}^*$ is determined, the
shower energy is estimated analytically.
We define a fiducial PMT annulus covering
$50\,\mathrm{m} \le R \le 200\,\mathrm{m}$ from the vertex. The inner
cut removes PMTs that are possibly saturated or too close to the vertex, while the outer cut matches the range of the PDF
tables.
The total predicted PE values from the shower at the reference energy
$E_{\rm ref} = 10^4\,\mathrm{GeV}$ summed over this fiducial annulus is,
\begin{equation}
    \mu_{\rm tot}^{\rm raw}
    \;=\;
    \sum_{i\in\mathrm{fiducial}}
    \Lambda(R_i)
    \cdot P_{\rm ang}(R_i,\cos\theta_i^*)
    \cdot w_{{\rm acc},i},
\label{eq:muraw}
\end{equation}
evaluated at the reconstructed direction.
The scaling factor that maps this reference prediction to the observed
total PE $Q_{\rm core}$ (summed over the same fiducial annulus) is,
\begin{equation}
    S \;=\; \frac{Q_{\rm core}}{\mu_{\rm tot}^{\rm raw}},
\label{eq:scaling}
\end{equation}
and the reconstructed energy follows from the linear proportionality
between deposited shower energy and total photon yield~\cite{icecube_energy},
\begin{equation}
    \log_{10}E_{\rm rec}
    \;=\;
    \log_{10}E_{\rm ref} + \log_{10}S
    \;=\;
    4.0 + \log_{10}\!\left(\frac{Q_{\rm core}}{\mu_{\rm tot}^{\rm raw}}\right).
\label{eq:erec}
\end{equation}
This analytic estimator requires no additional optimization and is
computed in a single pass directly over the fiducial PMTs once the direction is fixed. 
 
A residual bias in $E_{\rm rec}$ arises from the
threshold-limited detectors. Events near the lower boundary
of a reconstructed energy bin are preferentially drawn from the true
distribution above that boundary, causing $E_{\rm rec}$ to
systematically underestimate the true energy.
We correct for this with a linear calibration in logarithmic energy space.
Crucially, we bin the calibration profile by $\log_{10}E_{\rm true}$
rather than $\log_{10}E_{\rm rec}$: binning by the reconstructed
quantity would introduce regression dilution from the $\sim\!30\%$
energy resolution smearing.
The fitted formula is,
\begin{equation}
    \log_{10}E_{\rm cal}
    \;=\;
    0.982\;\cdot\;\log_{10}E_{\rm rec} \;+\; 0.087,
\label{eq:ecal}
\end{equation}
with slope close to unity confirming that the raw estimator is already
nearly unbiased in scale, and the small intercept correcting a residual
offset.

%% file: sec_perform_disc.tex
\section{Performance and Discussion}
\label{sec:performance}
 
\subsection{Angular Resolution}
\label{sec:angres}
 
\begin{figure}[htbp]
    \centering
    \includegraphics[width=0.45\linewidth,height=4.25cm]{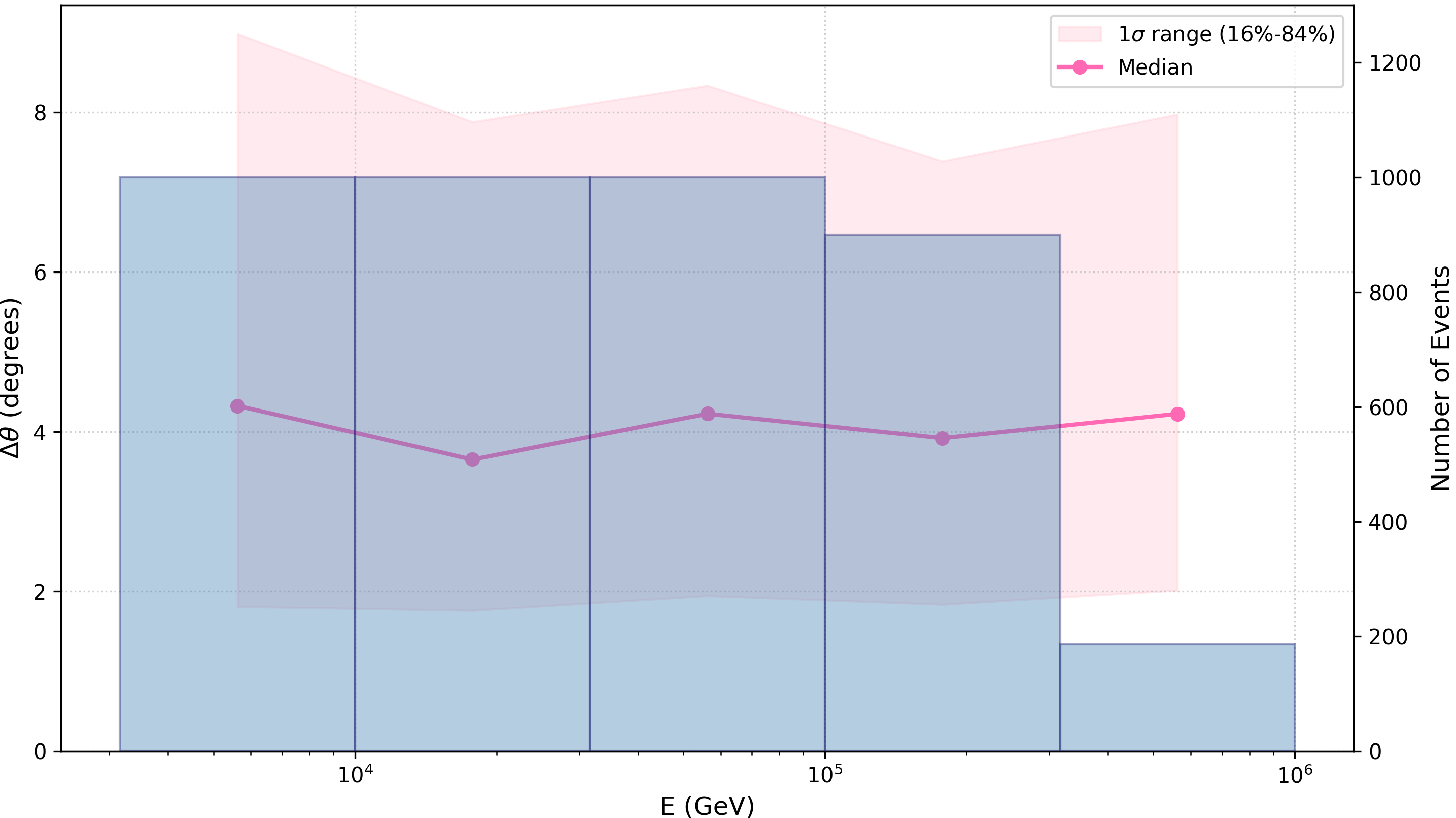}
    \includegraphics[width=0.45\linewidth]{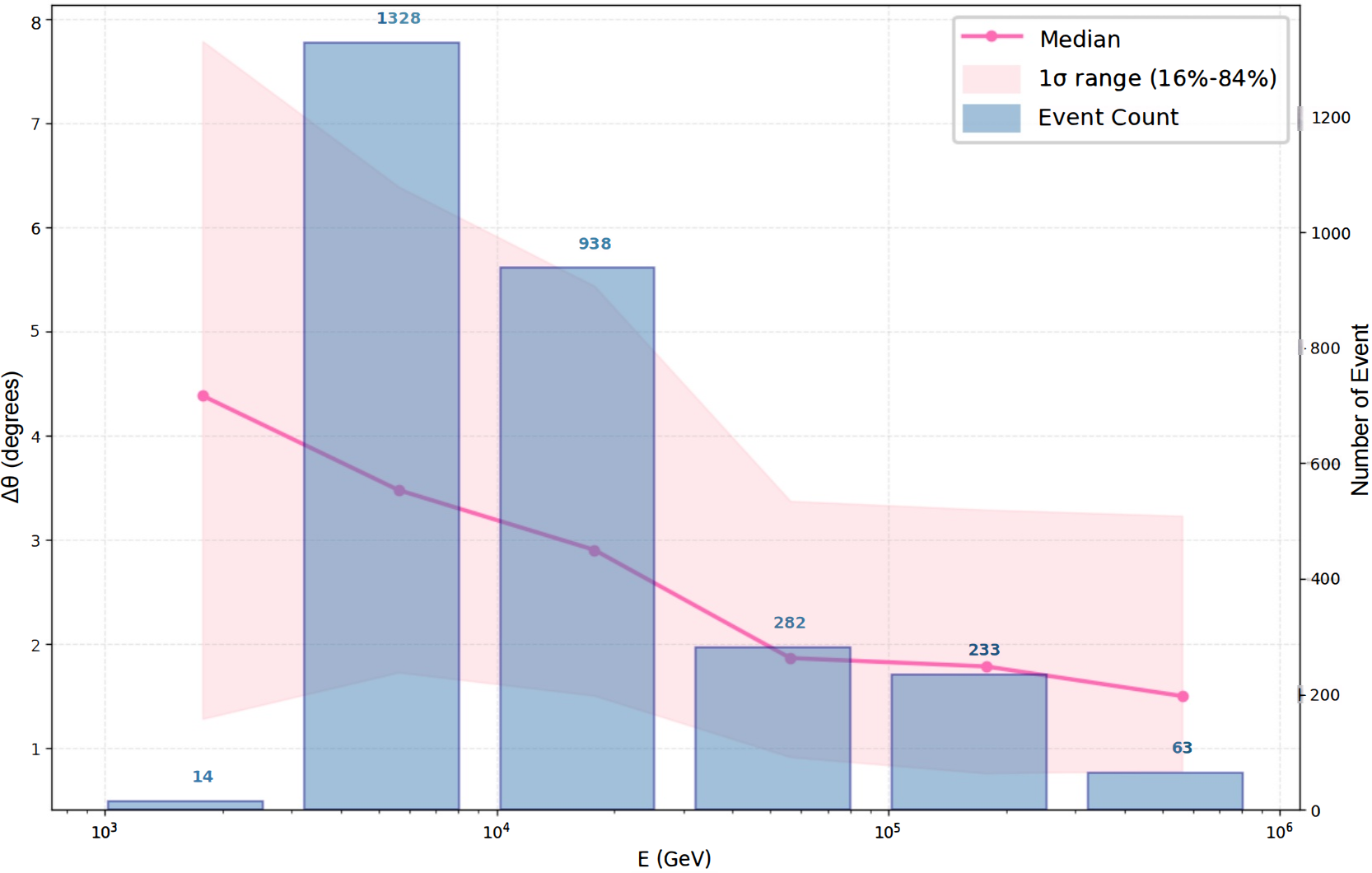}
    \caption{Angular resolution as a function of true neutrino energy. \textit{Left panel:} Traditional likelihood-based reconstruction. Points show the median angular error $\Psi$ per energy bin; the shaded band spans the 16th–84th percentile range. The bar chart (right axis) indicates the number of reconstructed events per bin. \textit{Right panel:} GNN-based reconstruction (from $1\,\mathrm{GeV}$ to $30\,\mathrm{TeV}$). The GNN consistently outperforms the likelihood-based method in the overlapping energy range.}
    \label{fig:ang_res}
\end{figure}

We characterize the directional performance using the spatial angle
between the reconstructed and true neutrino directions,
\begin{equation}
    \Psi \;=\; \arccos\!\left(\hat{\mathbf{d}}_{\rm rec}
                \cdot \hat{\mathbf{d}}_{\rm true}\right).
\label{eq:spaceangle}
\end{equation}
The angular resolution as a function of true neutrino energy is shown
in Figure~\ref{fig:ang_res}.
In the overlapping energy region $1$--$30\,\mathrm{TeV}$, the GNN
outperforms the likelihood-based method at every bin.

The GNN angular resolution is shown in Figure~\ref{fig:ang_res} (right panel)
as a function of true energy from $1\,\mathrm{GeV}$ to $30\,\mathrm{TeV}$
(in the range training statistics are sufficient).
It achieves a median error of $4.5^{\circ}$ at $1\,\mathrm{TeV}$,
improving rapidly to $2.8^{\circ}$ at $10\,\mathrm{TeV}$ and reaching
$1.8^{\circ}$ at $30\,\mathrm{TeV}$.
The 68th-percentile resolution (16th–84th range) narrows
monotonically from $2.8^{\circ}$--$7.8^{\circ}$ at $1\,\mathrm{TeV}$
to $1.0^{\circ}$--$2.5^{\circ}$ at $30\,\mathrm{TeV}$, indicating that
the GNN method produces a more peaked distribution with fewer large-angle
outliers.

Over the full energy range $1\,\mathrm{TeV}$--$1\,\mathrm{PeV}$,
the traditional MLE method achieves a median angular error of
$\Psi_{50} = 4.19^{\circ}$ with a 68th-percentile resolution of
$5.87^{\circ}$.
The $\Psi$ distribution is strongly peaked at small angles,
The median error improves from $4.91^{\circ}$ at $1$--$5\,\mathrm{TeV}$
to $3.97^{\circ}$ at $100\,\mathrm{TeV}$--$1\,\mathrm{PeV}$, consistent
with the expectation that higher-energy showers produce more Cherenkov
photons and thus provide stronger directional constraints.
A slight degradation is observed in the $50$--$100\,\mathrm{TeV}$ bin
($4.34^{\circ}$ median), which we attribute to the onset of shower
elongation effects that partially break the spherical symmetry assumed
in the PDF construction. 
The energy-binned angular performance is summarized in
Table~\ref{tab:angres}.
 
\begin{table}[ht]
\centering
\caption{Energy-binned angular performance for the traditional MLE
method.
$N$ is the number of reconstructed events. Median and 68\%~Q denote the median and 68th-percentile of the space
angle $\Psi$, respectively. While 16\% gives the 16th-percentile as a lower bound on resolution.}
\label{tab:angres}
\begin{tabular}{lcccc}
\hline
Energy range & $N$ & Median ($^{\circ}$) & 16\% ($^{\circ}$) & 68\% Q ($^{\circ}$) \\
\hline
$1$--$5\,\mathrm{TeV}$          & 3042  & 4.91 & 2.14 & 6.87 \\
$5$--$10\,\mathrm{TeV}$         & 1958  & 4.03 & 1.77 & 5.73 \\
$10$--$50\,\mathrm{TeV}$        & 4453  & 3.82 & 1.74 & 5.42 \\
$50$--$100\,\mathrm{TeV}$       & 547   & 4.34 & 2.15 & 6.07 \\
$100\,\mathrm{TeV}$--$1\,\mathrm{PeV}$ & 1086 & 3.97 & 1.87 & 5.29 \\
\hline
Total & 11086 & 4.19 & 1.89 & 5.87 \\
\hline
\end{tabular}
\end{table}

\begin{figure}[htbp]
    \centering
    \includegraphics[width=0.8\linewidth]{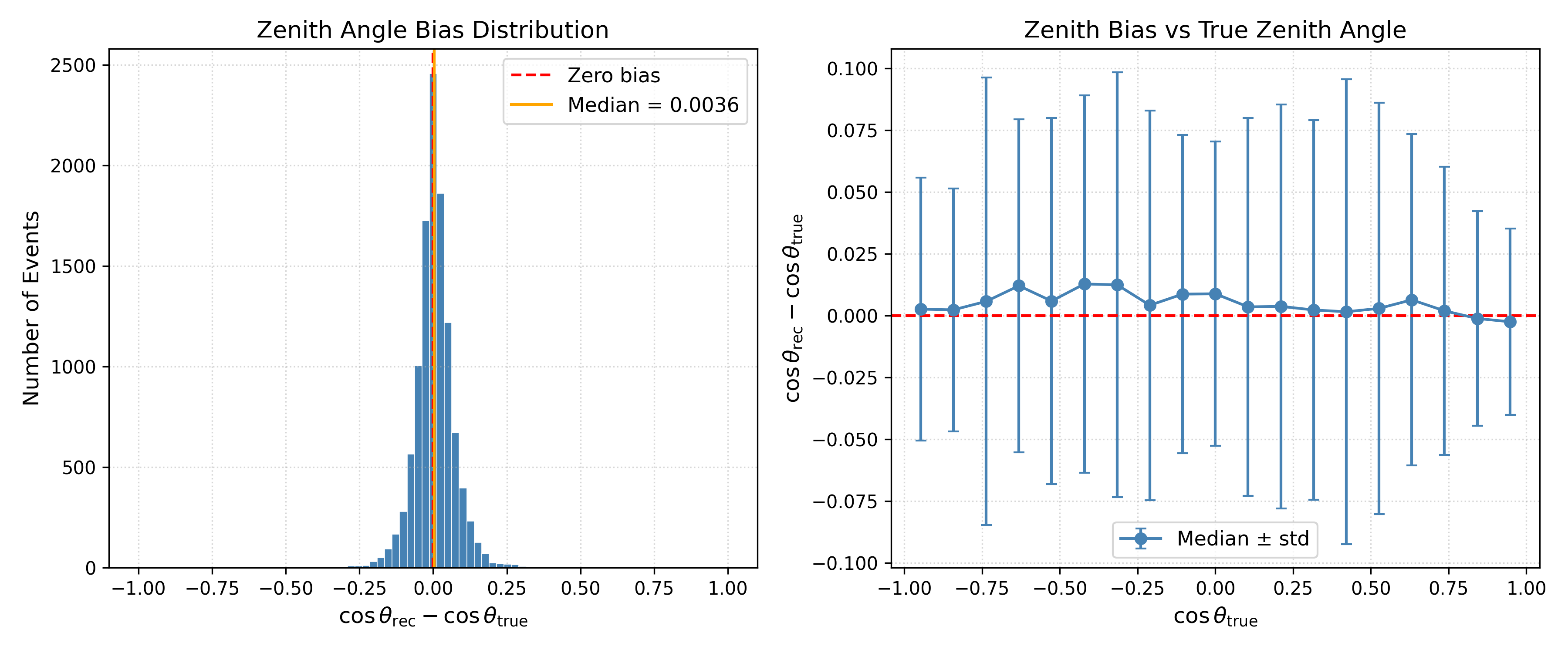}
    \caption{Zenith angle bias for the traditional MLE method. \textit{Left panel:} distribution of $\cos\theta_{\rm rec} - \cos\theta_{\rm true}$ over all reconstructed events; the distribution is centered at zero (median $= -0.003$) with no significant systematic shift. \textit{Right panel:} median bias as a function of $\cos\theta_{\rm true}$, showing no directional dependence across the full zenith range.}
    \label{fig:bias}
\end{figure}

 
We also checked for the MLE systematic bias in the directional reconstruction, and examine the distribution of $\cos\theta_{\rm rec} - \cos\theta_{\rm
true}$, where $\theta$ is the zenith angle.
As shown in Figure~\ref{fig:bias}, the distribution is centered
at zero, confirming the exclusion of a significant systematic shift in
the reconstructed zenith angle.
The distribution is also symmetric, demonstrating that the algorithm
does not preferentially reconstruct events toward or away from the
zenith.
The NEON array has an asymmetric Fibonacci
string layout and an anisotropic PMT acceptance model, both of which
could in principle introduce directional biases, while the MLE result is still reliable.
The right panel of Figure~\ref{fig:bias} shows the median bias
as a function of $\cos\theta_{\rm true}$, confirming that no
directional dependence is present across the full zenith range.
 
\subsection{Energy Reconstruction}
\label{sec:eres}
 
 
 
\begin{figure}[htbp]
    \centering
    \includegraphics[width=1.\linewidth]{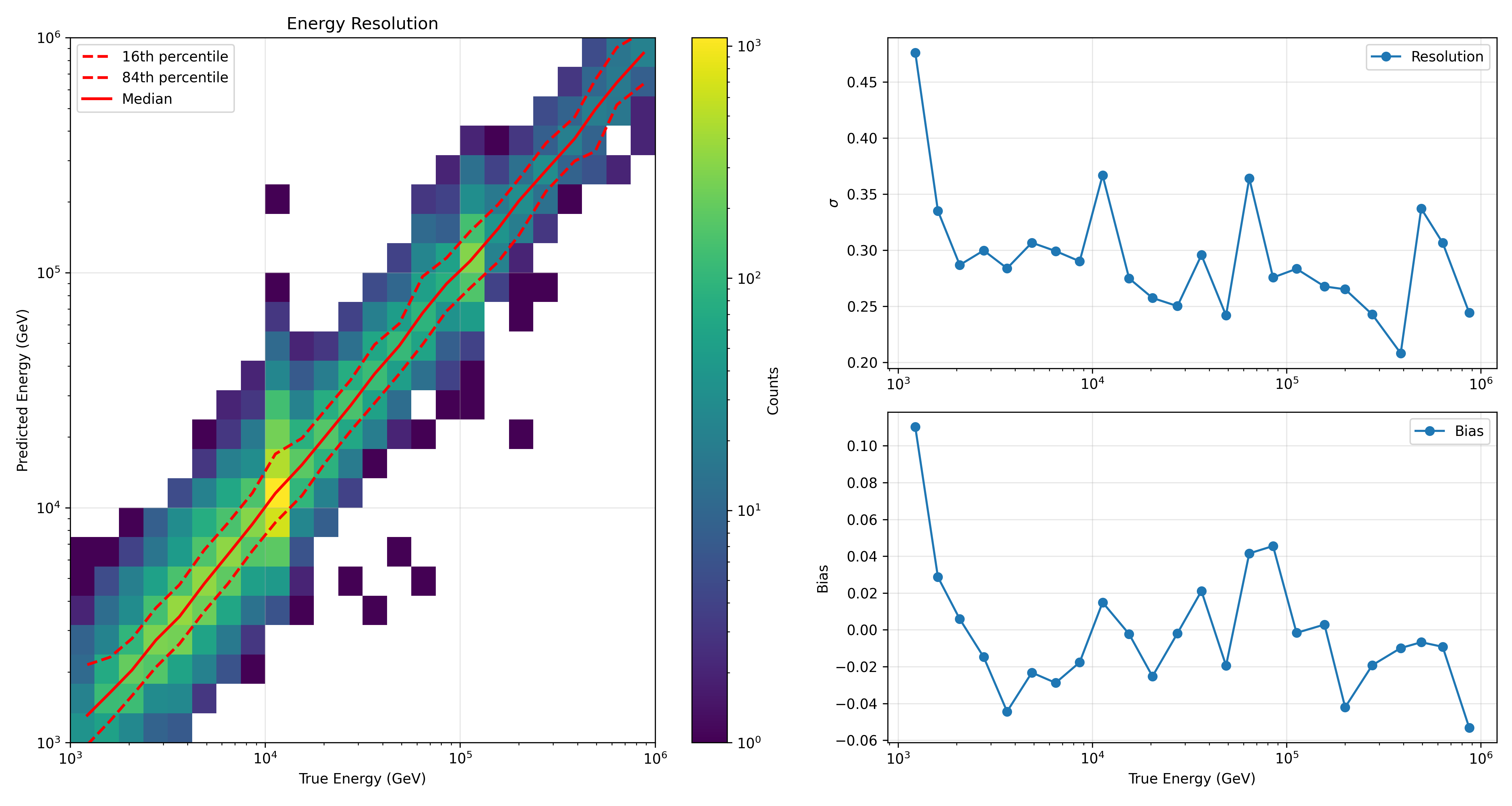}
  
    \vspace{4pt}
    
    \begin{minipage}[b]{0.5\linewidth}
        \centering \ \ 
        \includegraphics[width=\linewidth]{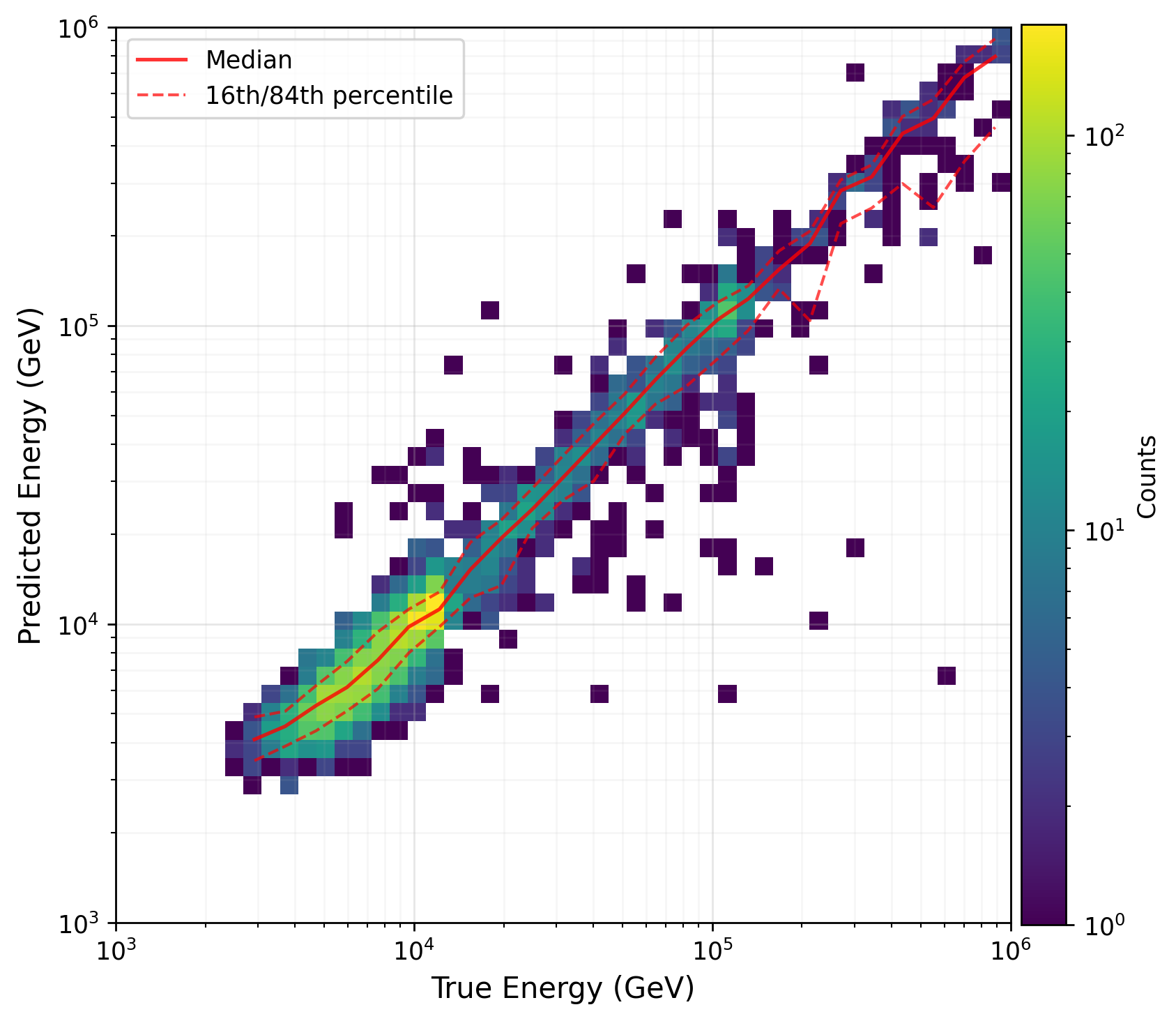}
        \label{fig:gnn_energy_dis}
    \end{minipage}
    \begin{minipage}[b]{0.49\linewidth}
        \centering
        \includegraphics[width=7.5cm,height=6.6cm]{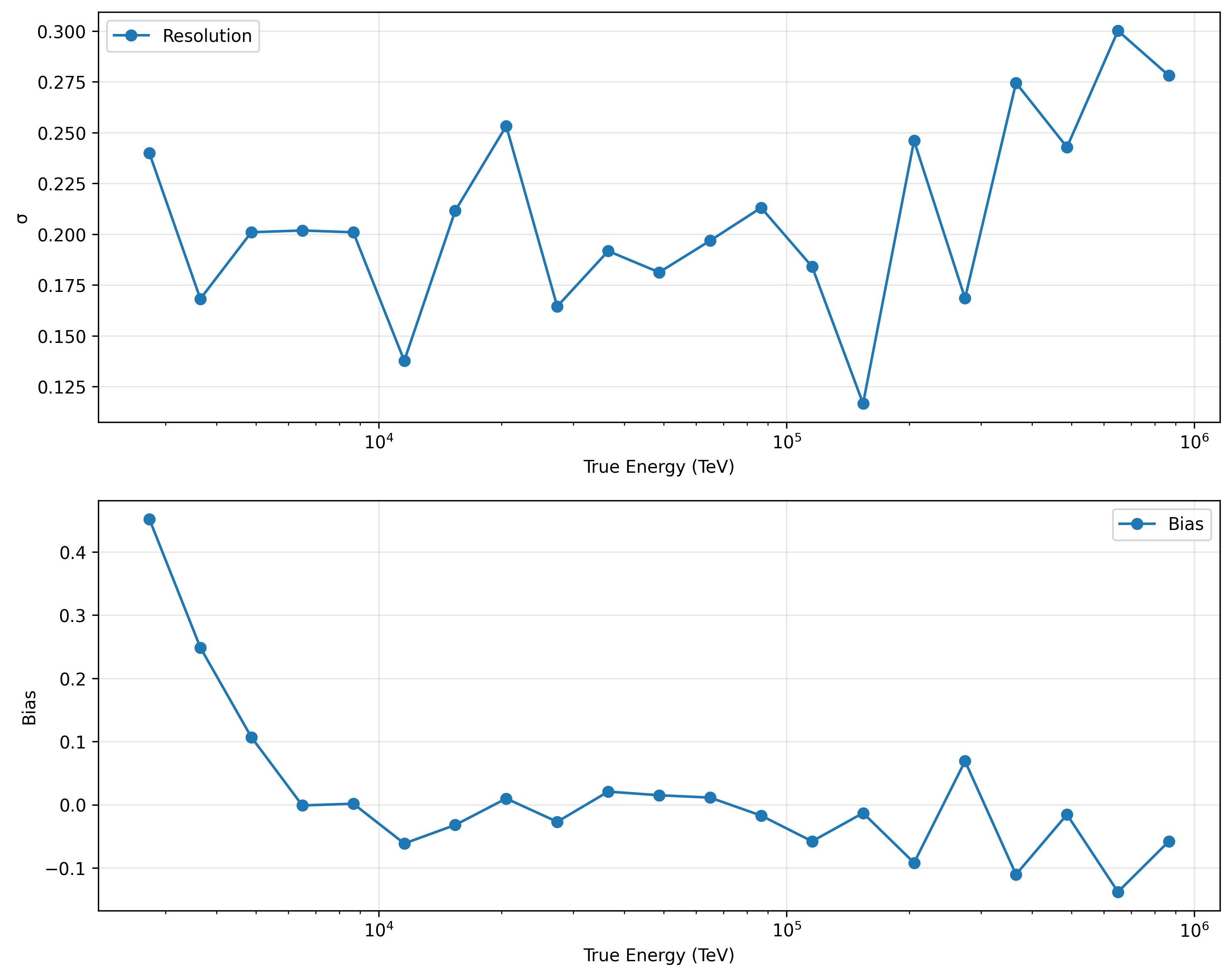}
        \label{fig:gnn_energy_res}
    \end{minipage}
    \caption{Energy reconstruction performance. \textit{Upper left panel:} Traditional likelihood-based method: 2D histogram of true versus reconstructed energy after calibration; red lines show the 16th, 50th, and 84th percentile profiles. \textit{Upper right panel:} Traditional likelihood-based method: energy resolution $\sigma_{E/E}$ as a function of true energy (upper subpanel) and median bias (lower subpanel). \textit{Lower left panel:} GNN-based method: scatter plot of true versus predicted energy. \textit{Lower right panel:} GNN-based method: energy resolution $\sigma$ (solid line) and median bias (dashed line) as functions of true energy.}
      \label{fig:erec}
\end{figure}

Figure \ref{fig:erec} shows the two-dimensional distribution of
$E_{\rm true}$ versus $E_{\rm cal}$, together with the 16th, 50th,
and 84th percentile profiles.
 
For the traditional method, the energy resolution, defined as
$\sigma_E/E = (E_{84} - E_{16})/(2\,E_{\rm true})$
evaluated in narrow logarithmic bins, ranges from approximately
$25\%$ to $37\%$ across the $1\,\mathrm{TeV}$--$1\,\mathrm{PeV}$
range.
The best performance of $\sim\!25\%$ is achieved around $50\,\mathrm{TeV}$,
where the number of triggered DOMs is large enough to provide good
charge statistics while high-energy saturation effects are still minor.
A mild increase is also observed at the highest energies ($\sim\!34\%$
above $400\,\mathrm{TeV}$), which we attribute to the small event
count in that regime rather than a genuine detector effect.
Nontheless, the energy-binned bias and resolution are summarized in
Table~\ref{tab:eres}.
The post-calibration median bias remains below $5\%$ in all energy
bins, confirming that the calibration effectively removes the
systematic offset across the full dynamic range.

For our GNN method directly trained from DOM hit patterns, 
Figure~\ref{fig:erec} shows the true versus
predicted energy for the GNN in the range $10$--$600\,\mathrm{TeV}$. The
points lie almost exactly on the diagonal, indicating a nearly perfect
linear response. Figure~\ref{fig:erec}  presents the energy
resolution $\sigma$ and median bias as functions of true energy from
$40$ to $300\,\mathrm{TeV}$. Over this range, the GNN achieves a
resolution between $0.14$ and $0.29$ (typically $\sim\!0.2$), substantially improving upon the likelihood-based
resolution of $\sim\!0.28$ in the same interval. The median bias remains
within $\pm15\%$ across most of the range.

\begin{table}[ht]
\centering
\caption{Energy-binned performance for the traditional MLE method
after calibration.
Bias is the median of $(E_{\rm cal} - E_{\rm true})/E_{\rm true}$;
$\sigma_E/E$ is the typical fine-bin resolution
$(E_{84}-E_{16})/(2\,E_{\rm true})$ read at the bin centre from
Figure~\ref{fig:erec}.}
\label{tab:eres}
\begin{tabular}{lccc}
\hline
Energy range & $N$ & Median bias & Typical $\sigma_E/E$ \\
\hline
$1$--$5\,\mathrm{TeV}$          & 3042 & $-0.015$ & $\sim\!0.33$ \\
$5$--$10\,\mathrm{TeV}$         & 1958 & $-0.022$ & $\sim\!0.30$ \\
$10$--$50\,\mathrm{TeV}$        & 4453 & $+0.007$ & $\sim\!0.27$ \\
$50$--$100\,\mathrm{TeV}$       & 547  & $+0.031$ & $\sim\!0.25$ \\
\end{tabular}

\end{table}

\subsection{Effective Area and Point-source Sensitivity}
\label{sec:aeff}

\begin{figure}[htbp]
    \centering
    \includegraphics[width=0.6\linewidth]{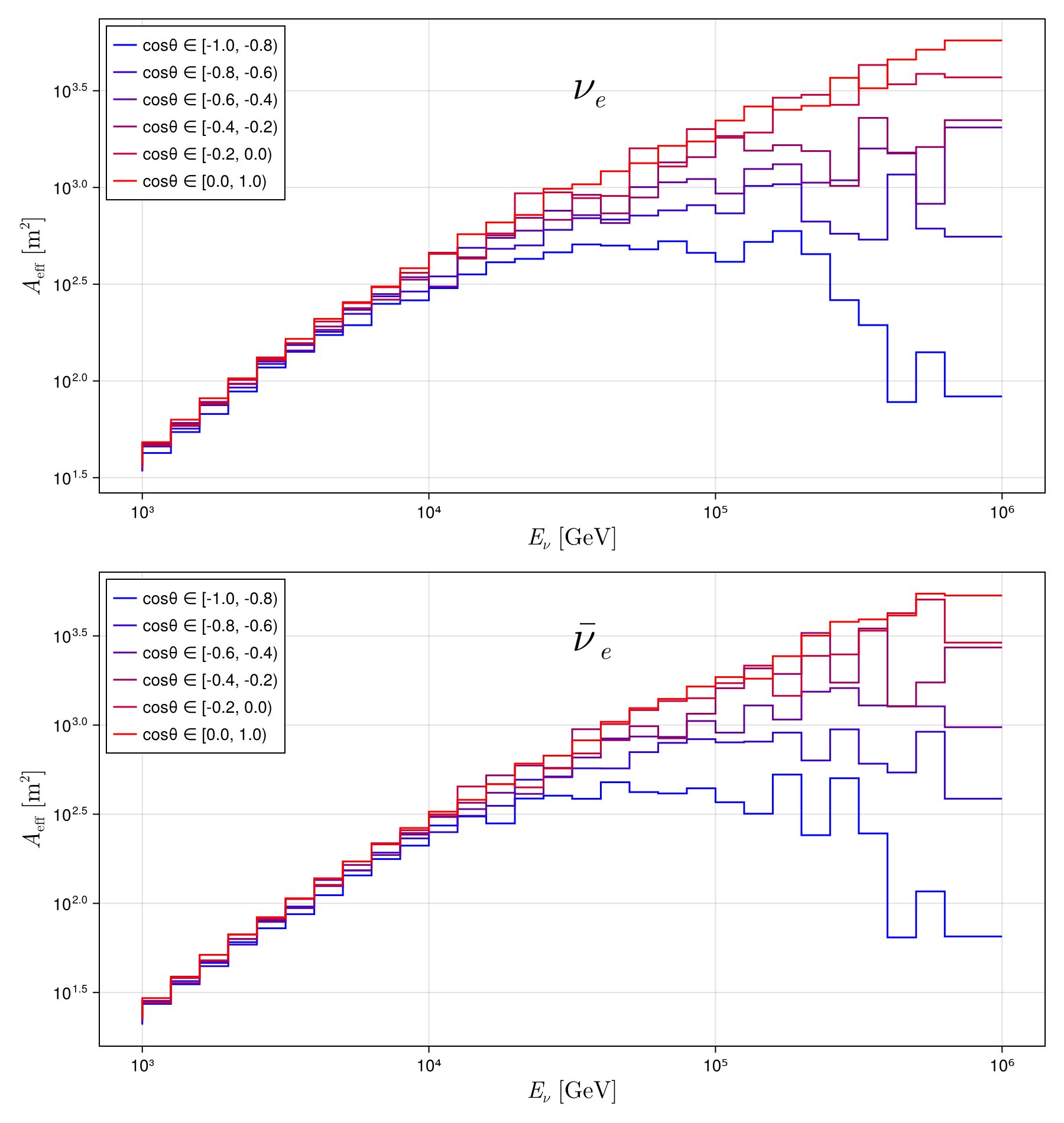}
    \caption{Effective area of NEON for shower-like events as a function of neutrino energy. Upper (lower) panel: $\nu_e$ ($\bar\nu_e$). Different colours correspond to six bins of $\cos\theta$ from $-1$ (up-going, blue) to $1$ (down-going, red). The effective area includes the trigger, hit selection, and reconstruction efficiency.}
    \label{fig:effective_area}
\end{figure}

The effective area of the experiment is computed from the simulated shower-like events together with the reconstruction efficiency of this work. Figure~\ref{fig:effective_area} shows the reconstructed effective area $A_{\rm eff}(E_{\nu},\cos\theta)$ for $\nu_e$ (upper panel) and $\bar\nu_e$ (lower panel), binned into six $\cos\theta$ intervals. The effective area grows from about $3\times10^{2}\,\mathrm{m^2}$ at 10\,TeV to roughly $10^{3}$--$2\times10^{3}\,\mathrm{m^2}$ at 100\,TeV, and its energy dependence differs among directions. For deeply up-going directions ($\cos\theta \lesssim -0.6$) the growth weakens at high energies, while only near-horizontal and down-going directions ($\cos\theta \gtrsim 0$) keep increasing with energy up to the highest simulated energies.

\begin{figure}[htbp]
    \centering
    \includegraphics[width=0.6\linewidth]{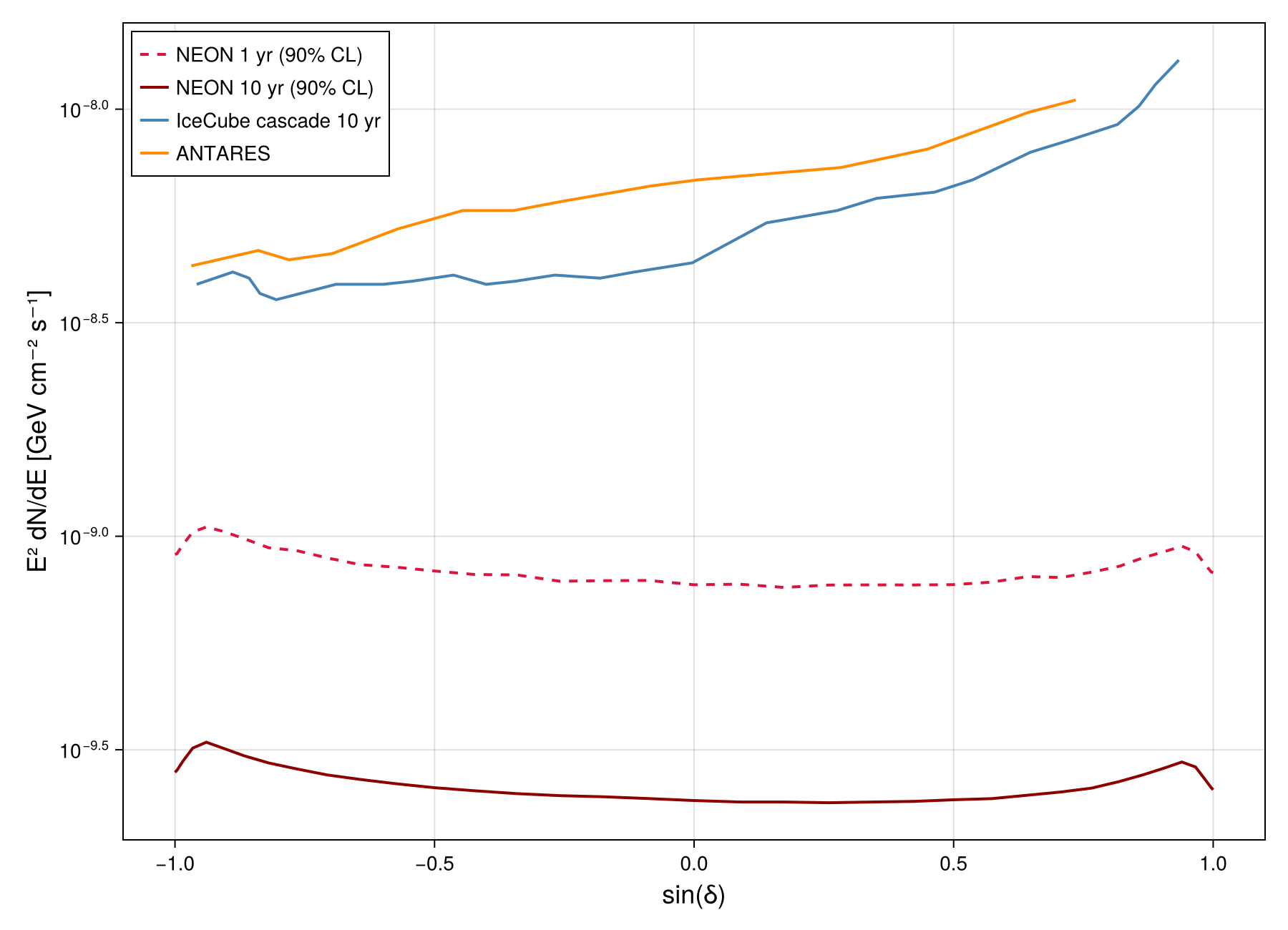}
    \caption{Expected 90\% CL sensitivity to point sources with an $E^{-2}$ spectrum, quoted as $E^2\,\mathrm{d}N/\mathrm{d}E$ at 100\,TeV as a function of the source declination $\sin\delta$. The dashed and solid crimson/darkred curves show NEON for one and ten years of operation, respectively, using events above 100\,TeV; the steel-blue and orange curves show the ten-year cascade sensitivities of IceCube and ANTARES~\cite{IceCube:2023gp}.}
    \label{fig:sensitivity}
\end{figure}

Based on the effective area and the reconstruction performance presented above, we estimate the sensitivity of NEON to point sources assuming an $E^{-2}$ neutrino spectrum. The signal search window is set from the angular-error distribution of the GNN reconstruction (quantiles of the error distribution rather than a Gaussian assumption), the atmospheric neutrino background is taken from the Honda flux model, and the isotropic astrophysical neutrino flux measured by IceCube with cascade events~\cite{IceCube:2020cascade} is included as an additional background component. The 90\% confidence level limits are obtained with the Feldman--Cousins method. To match the reference energy of the comparison analyses, only events above 100\,TeV are retained, and the resulting flux is quoted at $E=100\,\mathrm{TeV}$. Figure~\ref{fig:sensitivity} shows the expected $E^2\,\mathrm{d}N/\mathrm{d}E$ at 100\,TeV as a function of the source declination $\sin\delta$ for one and ten years of operation, together with the ten-year cascade sensitivities of IceCube and ANTARES~\cite{IceCube:2023gp}.

%% file: sec_conclu.tex
\section{Conclusion}
\label{sec:conclu}

In this work, we have developed and evaluated the  reconstruction framework dedicated to high-energy shower-like neutrino events for NEON. Accurate reconstruction of cascade events in a deep-sea Cherenkov telescope is challenging due to seawater optical dispersion, irregular detector geometry, and severe $^{40}\text{K}$ ambient background noise. To address these, we implemented and compared two complementary methodologies, a physics-driven MLE pipeline and a data-driven hierarchical GNN.

The traditional MLE framework combines preprocessing with detailed physical calibrations. An isochronic spatial-temporal hit-cleaning algorithm successfully rejects over 98\% of uncorrelated $^{40}\text{K}$ background hits. Vertex reconstruction employing a robust M-estimator on arrival-time residuals delivers sub-meter spatial precision, mitigating large scattering-induced tail fluctuations. Crucially, the directional and energy likelihood formulations incorporate precise physical corrections, including PMT directional angular acceptance, charge-dependent time slewing, and a finite line-source model that accounts for the longitudinal extension of electromagnetic and hadronic cascades. Across the energy range from $1\text{ TeV}$ to $1\text{ PeV}$, the MLE method achieves an overall median angular resolution of $4.19^\circ$ and an energy resolution of $25\%$--$37\%$, with residual systematic biases constrained within $0.05$ in $\log_{10}(E_{\text{reco}}/E_{\text{true}})$.

To exploit complex spatiotemporal topologies and enhance computational efficiency, we introduced a two-stage GNN. By decoupling local intra-DOM PMT features from distance-weighted inter-DOM graph convolutions, the network captures fine-grained Cherenkov wavefront characteristics. In the low-to-intermediate energy regime ($10\text{ TeV} \lesssim E_\nu \lesssim 300\text{ TeV}$), where sparse hit multiplicities limit likelihood optimization, the GNN demonstrates remarkable superiority, achieving a median angular resolution of $1.8^\circ$ at $30\text{ TeV}$ and an energy resolution of $\sim 20\%$. Furthermore, the GNN achieves an inference throughput of several milliseconds per event, making it an ideal candidate for real-time online alert systems.

Based on these reconstruction pipelines, we evaluated the scientific performance of the full NEON array. The electron neutrino effective area reaches $\sim 1\text{ m}^2$ at 
$100\text{ TeV}$ and exceeds $10\text{ m}^2$ in the PeV regime. Point-source sensitivity analysis confirms that the improved cascade angular resolution substantially reduces the background contamination, boosting NEON's discovery potential for southern-sky astrophysical sources and diffuse flux spectrum.

Nevertheless, several improvements will further advance NEON's reconstruction capabilities. Future iterations will integrate in-situ deep-sea optical calibrations, including depth- and season-dependent absorption and scattering profiles obtained from ongoing sea trials, into both the likelihood PDF tables and GNN training datasets to mitigate environmental systematic uncertainties. Moreover, the hierarchical graph framework can be extended toward a unified multi-task architecture, enabling simultaneous event topological classification (tracks, cascades, and double-bangs) alongside full-parameter reconstruction. Finally, leveraging the millisecond-scale inference speed of the GNN, we plan to deploy optimized deep learning models onto onboard GPU or FPGA edge-computing platforms, establishing a low-latency ($\lesssim 1\text{ s}$) online alert pipeline to facilitate rapid multi-messenger astronomical follow-ups.

In summary, this work establishes a rigorous reconstruction baseline for shower-like events in NEON, showing that the synergy of traditional likelihood methods and deep learning architectures provides a powerful pathway toward high-precision neutrino astronomy in deep-sea environments.

%% file: appendix.tex
\appendix
\section{Appendix for the MLE Method}
\label{app:mle}

\subsection{Pre-computed PDF Tables}
\label{sec:pdftables}

We pre-compute three lookup tables from the full simulation sample over the fiducial range $15\,\mathrm{m} \le R \le 200\,\mathrm{m}$, following the approach used in deep-sea and ice-based neutrino telescopes~\cite{icecube_energy,2017ICRC...35..950M,KM3NeT:2024awa},

\begin{itemize}
    \item $\Lambda(R)$: the expected total photon yield in PE, encoding radial light attenuation in seawater.
    \item $P_{\rm ang}(R,\cos\theta)$: the normalized angular probability of a photon hit, describing the Cherenkov cone geometry.
    \item $P_{\rm time}(R,\cos\theta,t_{\rm res})$: the time-residual probability distribution, describing the spread in photon arrival times due to scattering.
\end{itemize}

The photon yield and angular distribution depend on the distance $R$ from the shower vertex and the angle $\cos\theta$ between the shower axis and the vertex-to-PMT direction. These are evaluated at runtime by linear interpolation from the tables.

Since the Cherenkov photon yield scales linearly with shower energy, we marginalize the energy out of $\Lambda$ and $P_{\rm ang}$ by reweighting each simulated hit by $w_E = E_{\rm ref}/E$ before filling the histograms, where $E_{\rm ref} = 10^4\,\mathrm{GeV}$ is the reference energy. This rescales each event's contribution to the equivalent at $E_{\rm ref}$, so the accumulated tables describe a single reference energy regardless of the original simulation sample. The direction search then operates over $(d_x, d_y, d_z)$ only, and energy is recovered analytically afterwards (Section~\ref{sec:energy}).
Gaussian smoothing is applied to $P_{\rm ang}$ and $P_{\rm time}$ after filling to suppress statistical fluctuations. $P_{\rm time}$ is normalized to unit integral over $t_{\rm res}$ in each $(R,\cos\theta)$ cell.

\subsection{Physical Calibrations}
\label{sec:calib}
 
Two hardware-level effects must be corrected before the predicted and
observed photon distributions can be meaningfully compared.
 
\begin{figure}[htbp]
    \centering
    \includegraphics[width=0.6\linewidth]{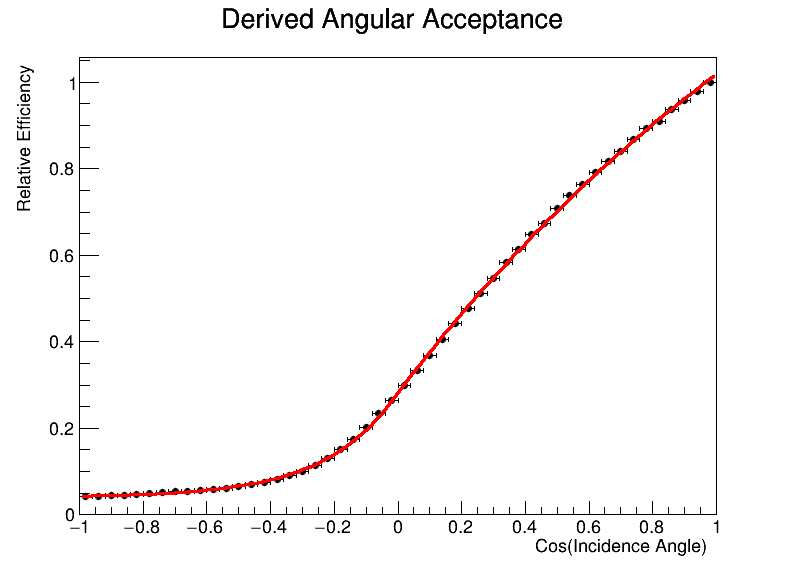}
    \caption{PMT angular acceptance as a function of photon incident angle $\cos\eta$. Points are derived from simulation; the red curve is the fitted sigmoid model (Eq.~\ref{eq:acceptance}).}
    \label{fig:acc_fit}
\end{figure}

\paragraph{PMT Angular Acceptance.}
The detection efficiency of a PMT depends strongly on the incident angle
$\eta$ of the photon relative to the PMT normal.
We model this response with a sigmoid function fitted to simulation data,
\begin{equation}
    w_{\rm acc}(\cos\eta)
    \;=\;
    p_0 \;+\; \frac{p_1}{1+\exp\!\left(-p_2(\cos\eta - p_3)\right)},
\label{eq:acceptance}
\end{equation}
with calibrated constants $p_0 = 0.03636$, $p_1 = 0.9413$,
$p_2 = 4.397$, $p_3 = 0.2593$ obtained from a least-squares fit
to the simulated acceptance curve.
The sigmoid form was chosen because it naturally reproduces the
observed behaviour: near-zero efficiency at backward angles
($\cos\eta \approx -1$) transitioning smoothly to near-unity
efficiency at normal incidence ($\cos\eta \approx 1$), as shown in
as shown in Figure~\ref{fig:acc_fit}.
The acceptance weight $w_{{\rm acc},i}$ is pre-computed once per PMT
after the vertex position is known.
 
\begin{figure}[htbp]
    \centering
    \includegraphics[width=0.6\linewidth]{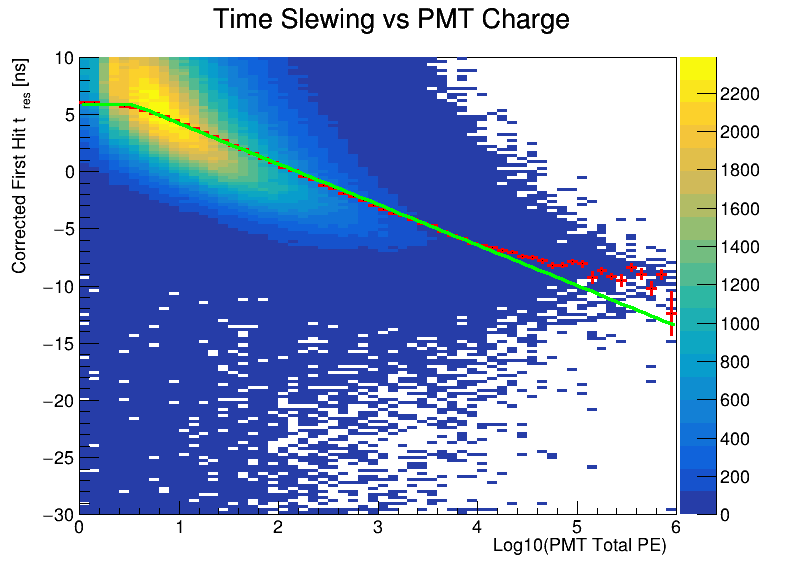}
    \caption{Time slewing statistics. Measured time shift as a function of $\log_{10}(q/\mathrm{PE})$. The red error bars and green line show the estimated and fitted time shifts.}
    \label{fig:slewing}
\end{figure}

\paragraph{Time Slewing Correction.}
We discover a PE-dependent time drift relative to the ideal Cherenkov light front, likely caused by delayed photons from multiple scattering, and we here call it time slewing effect. Figure \ref{fig:slewing} shows the statistical correlation, and we correct for it with an empirical fit to the PE--time relation:
\begin{equation}
    t_{\rm corr} \;=\; t \;-\; \Delta t_{\rm slew}(\log_{10} q),
\label{eq:slewing}
\end{equation}
where $q$ is the hit PE and
\begin{equation}
    \Delta t_{\rm slew}
    \;=\;
    \begin{cases}
        0,
        & \log_{10} q \le 0.541,\\[4pt]
        -3.579\,\times\,(\log_{10} q_{\rm eff} - 0.541),
        & \log_{10} q > 0.541,
    \end{cases}
\label{eq:slewcurve}
\end{equation}
with $q_{\rm eff} = \min(q,\,10^{4.5193})$ to avoid extrapolation
into the saturation regime.
The threshold $0.541$ and slope $-3.579$ were determined by a linear
fit to the measured time-advance profile shown in
Figure~\ref{fig:slewing}. The saturation cap $10^{4.5193}$ was set at
the largest charge bin with sufficient statistics in the calibration
sample.
We note that the correction is applied per individual hit rather than per PMT.
 
\subsection{Line Source Model for Electromagnetic Shower}
\label{app:lsm}

 
An electromagnetic (EM) shower after a neutrino
interaction develops along the direction of the primary lepton.
The shower begins at the interaction vertex and grows along
$\hat{\mathbf{d}}$, reaching its maximum particle multiplicity at a
radiation depth of $X_{\max} \sim \ln(E/E_c)$ radiation lengths, where
$E_c \approx 73\,\mathrm{MeV}$ is the critical energy in water.
Therefore, Cherenkov photons are actually emitted not from a single point but from
a spatial extent along the shower axis that depends on energy.
In the energy range of interest ($1\,\mathrm{TeV}$--$1\,\mathrm{PeV}$),
for instance, $X_{\max}$ at $1\,\mathrm{TeV}$ corresponds
to roughly $4$--$5\,\mathrm{m}$ of physical depth.

We note that such an effect is significant especially for hits near  the vertex region.   
Ignoring this effect introduces a systematic early-time bias in the time residual $t_{\rm res}$, shifting the peak of the $P_{\rm time}$ distribution and degrading the temporal likelihood. 
Hence, incorporating the line-source geometry of shower into the temporal term promote the direction reconstruction performance what the purely spatial PE distribution provides. 


We therefore adopt a fixed effective shower length of
$L_{\rm shower} = 4.5\,\mathrm{m}$, which represents the characteristic
physical extent of a $1$--$10\,\mathrm{TeV}$ EM cascade in seawater and
is consistent with shower-profile parameterisations used in the
literature~\cite{icecube_energy,2017ICRC...35..950M}.
The value $L_{\rm shower} = 4.5\,\mathrm{m}$ was determined empirically
by scanning values from $1\,\mathrm{m}$ to $20\,\mathrm{m}$ on a
held-out validation sample and selecting the length that minimized the
median angular error $\Psi_{50}$ across the full energy range.
The optimum is broad and relatively flat between $3\,\mathrm{m}$ and
$8\,\mathrm{m}$, consistent with the physical shower-length range, and
the performance degrades significantly only for $L_{\rm shower} < 1\,\mathrm{m}$
(approaching the point-source limit) or $L_{\rm shower} > 15\,\mathrm{m}$
(where the model overestimates the shower extent and introduces spurious
time corrections for geometrically disfavoured PMTs).
The corresponding geometric time advance $\delta t_i$ relative to
a point source at the vertex is then computed from the distances
traveled in vacuum and in water.
For hit $j$ at PMT $i$ with individual charge $q_j$,
the purified time residual is:
\begin{equation}
    t'_{{\rm res},j}
    \;=\;
    t_{{\rm res},j} \;-\; \delta t_i \;-\; \Delta t_{{\rm slew},j}.
\label{eq:purified}
\end{equation}